\documentclass[11pt, a4paper]{article}

\usepackage{marginnote}
\usepackage{tikz}
\usepackage{natbib}
\usepackage{bm}
\usepackage{amsmath}
\usepackage{tabularx}
\usepackage{amssymb}
\usepackage{amsthm}
\usepackage{booktabs}
\usepackage{multirow}
\usepackage{graphicx}
\usepackage{mathdots}
\usepackage{pdfpages}
\usepackage{url}
\usepackage[left = 2.5cm, right = 2.5cm, top = 2cm, bottom = 3cm]{geometry}

\author{Ioannis Kosmidis \\ Department of Statistical Science,
  University College London \\
  Gower Street, London, WC1E 6BT, United Kingdom \\
  \texttt{i.kosmidis@ucl.ac.uk}
  \bigskip \\
  and \bigskip \\
  Louis Passfield \\ Endurance Research Group, \\ School of Sport and
  Exercise Sciences, University of Kent, \\Chatham Maritime, Chatham,
  Kent, ME4 4AG,
  United Kingdom \\
  \texttt{l.passfield@kent.ac.uk}}

\title{Linking the performance of endurance runners to training and
  physiological effects via multi-resolution elastic net}

\begin{document}
\maketitle

\begin{abstract}
  A multiplicative effects model is introduced for the identification
  of the factors that are influential to the performance of
  highly-trained endurance runners. The model extends the established
  power-law relationship between performance times and distances by
  taking into account the effect of the physiological status of the
  runners, and training effects extracted from GPS records collected
  over the course of a year.  In order to incorporate information on
  the runners' training into the model, the concept of the training
  distribution profile is introduced and its ability to capture the
  characteristics of the training session is discussed. The covariates
  that are relevant to runner performance as response are identified
  using a procedure termed multi-resolution elastic
  net. Multi-resolution elastic net allows the simultaneous
  identification of scalar covariates and of intervals on the domain
  of one or more functional covariates that are most influential for
  the response. The results identify a contiguous group of speed
  intervals between 5.3 to 5.7 m$\cdot$s$^{-1}$ as influential for the
  improvement of running performance and extend established
  relationships between physiological status and runner
  performance. Another outcome of multi-resolution elastic net is a
  predictive equation for performance based on the minimization of the
  mean squared prediction error on a test data set across resolutions.
\end{abstract}

\section{Introduction}
Competitive runners focus upon training effectively in order to
enhance their fitness and performance. Yet despite many advances in
the scientific evaluation of responses to training, the prescription
and effectiveness assessment of training programmes relies upon the
intuition and experience of runners and their coaches.

The early attempts to model the effects of training on performance
were pioneered by \citet{banister:75} who proposed an impulse-response
model (see, also, \citealt{busso:03} for a more recent application of
the \citealt{banister:75} methodology). Banister and colleagues
quantified training impulse as a single model input using arbitrary
units, where the response is modelled as a change in performance that
varies according to the training input. However, runners cannot use
this model to inform their training as it requires frequent
performance trials that interfere with their training
programme. Several fundamental issues have also been highlighted by
\citet{busso:06} and \citet{jobson:09}. Specifically, the model needs
input training data to be aggregated according to an assumed
biological or physiological model. This biological or physiological
model provides an abstraction of the complex relation between training
input and response but has never been validated for this
purpose. Critically, the necessary data aggregation restricts the use
of the model to evaluating only programmes comprised of identical
training sessions. Ultimately, the limitations of the model are such
that its derived parameter estimates are not generalizable beyond the
training session or participant studied.

A superior approach to modelling training would be to characterise its
effects on race performance, i.e. as the time to cover a specified
race distance. The relationship between \emph{best} race performances
over varying race distances has been found to follow a standard
exponential curve over a wide range of values. \citet{kennely:06} was
the first to illustrate that the power law is a very good fit to this
relation by plotting world best performances for men running from 100
m to 100 km, with an average deviation of only 4.3\%. The more recent
studies of \citet{katz:99} and \citet{savaglio:2000} have also
confirmed that a power law fits well the relationship of best running
performance and race distance. These models, however, do not account
for the effect that training or the physiological status of the runner
has on their performance, and hence they are overly simplistic into
describing how the performance of individual runners varies
accordingly.

With recent developments in training technology runners can now use
GPS devices to record their training and races with a level of
accuracy and detail that was previously inconceivable. The capability
to obtain large volumes of training data from runners presents the
opportunity for new insights into the links between training and
performance by removing the need to use the limited impulse response
model and by extending well-established parametric relationships for
performance. That is the primary aim of the present study. The present
study investigates the effects of training and physiological factors
on the performance of highly-trained runners' competing in distances
from 800 metres to marathon (endurance runners, for short). A
secondary aim is to produce a predictive equation for race
performance. The available data are from a year-long observational
study of 14 endurance runners, which produced detailed GPS records of
their training, their physiological status and their best performance
in standardised field tests \citep{galbraith:14}.

The contribution of the present work is two-fold. Firstly, a
multiplicative effects model for the performance of endurance runners
is constructed, which extends the well-studied power-law relationship
between runners' performance times and distances by also taking into
account the physiological status of the runner and information on the
runners' training. Particularly, in order to capture information on
the runners' training, the concept of the training distribution
profile is introduced. The resulting model involves performance as a
scalar response, a group of associated scalar covariates and the
training distribution profile, which is a functional covariate. In
order to simultaneously identify the speed intervals on the domain of
the training distribution profile and the scalar physiological
covariates that are important for explaining performance, we introduce
a procedure termed multi-resolution elastic net. Multi-resolution
elastic net proceeds by combining the partitioning of the domain of
functional covariates in an increasing number of intervals with the
elastic net of \citet{zou:05}, and results in predictive equations
that involve only interval summaries of the functional covariates. We
argue that the usefulness of multi-resolution elastic net extends
beyond the present study. The supplementary material provides a
reproducible analysis of a popular data set in functional regression
analysis using multi-resolution elastic net.

The structure of the manuscript is as follows. Section~\ref{studydata}
provides a description of the available data, along with an overview
of the protocols used for data collection, the performance assessments
and the collection of physiological status information from laboratory
tests. Section~\ref{modelling} works towards setting up a general
multiplicative effects model for performance that decomposes the
performance times into effects due to race distance, physiological
status, training and other effects. Section~\ref{estimation} describes
multi-resolution elastic net and its use for estimating the parameters
of the multiplicative model. The outcomes of the modelling exercise
are described in Section~\ref{results} and confirm and extend
established relationships between physiological status and runner
performance, and, importantly, identify a contiguous group of speed
intervals between 5.3 to 5.7 m$\cdot$s$^{-1}$ as influential for
performance. A predictive equation for performance is also provided by
the minimization of the estimated mean squared prediction error
estimated from a test set across resolutions. The manuscript concludes
with Section~\ref{discussion} which provides discussion and directions
for further work.

\section{Data}
\label{studydata}

\subsection{Participants}
The available data were gathered as part of the study in
\citep{galbraith:14} which examines changes in laboratory and field
test running performances of highly-trained endurance runners.
The study involved the observation of 14 competitive endurance
runners, who had a minimum of 8 years experience of running training,
and competition experience in race distances ranging from 800m to
marathon.
All participants provided written informed consent for
this study which had local ethics committee approval from the
University of Kent, Chatham Maritime, United Kingdom.
More details on the participants, the study design
and data collection protocols are provided in \citet{galbraith:14},
but a brief account is also provided below.

\begin{figure}[t!]
  \begin{center}
    \includegraphics[width = \textwidth]{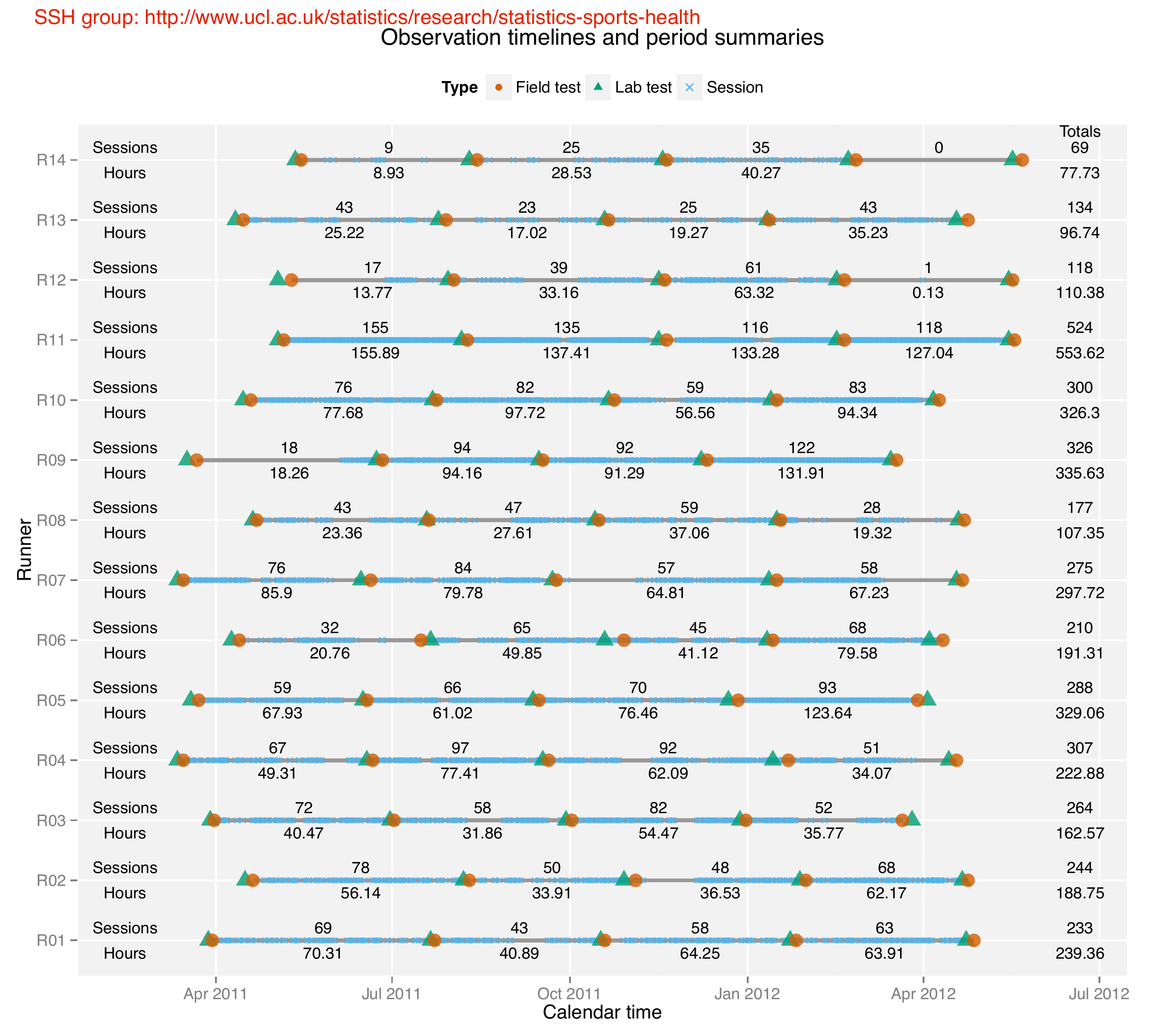}
  \end{center}
  \caption{The observation timelines for the runners with the events
    of the study (lab tests, field tests and the recorded sessions for
    each runner) on a calendar scale. The number of recorded sessions
    and number of training hours within each training period and the
    corresponding totals are shown above, below and on the right of
    each timeline. In order to preserve anonymity the runners are
    referred to as R01, R02, $\ldots$, R14.}
  \label{study}
\end{figure}

\subsection{Data collection}
\label{collection}
The data was gathered by observing the training of the participating
runners for a year. On commencing the study each runner was supplied
with a wrist-worn GPS device (Forerunner$^\copyright$ 310XT, Garmin
International Inc. Kansas, USA) and instructed in its use according to
the manufacturer's guidelines. The runners were asked to use the GPS
device to record their training time and distance throughout every
session and race in the observation period. The study did not involve
any direct manipulation of the runners' training programmes and the
runners regularly downloaded the data from their GPS devices and sent
the resulting files to the lead scientist.

In addition to their training, each runner completed 5 laboratory
tests and 9 track-based field tests at regular intervals throughout
the study. The laboratory tests were used to measure traditional
physiological status determinants of running performance, i.e. running
economy, OBLA (on-set blood lactate accumulation), and
$\dot{V}O_{2max}$ (a measure of maximum oxygen consumption). The
track-based field tests were conducted to measure the runners' best
performance over distances of $D_1 = 1200$ metres, $D_2 = 2400$ metres
and $D_3 = 3600$ metres.

For the purposes of the current study, only the field tests that
occurred within a few days of the laboratory tests are considered,
i.e. 5 out of the 9 field tests for each runner. The complete
observation period for each runner was set to the interval between and
including the runner's first and last field test.

Prior to all laboratory and field tests, careful standardisation
ensured that each test was completed under conditions where the time
of day, prior exercise, diet, hydration and warm up were specified and
controlled. The runners were also familiarised with all laboratory and
field tests before commencing the study.

\subsection{Extraction of training sessions and speed profiles}

\label{protocol}
The raw GPS data consists of 2,499,894 timestamped measurements of
cumulative distances calculated using latitude and longitude
information for the complete observation period for each runner. Those
raw observations were used to identify 3,469 distinct training
sessions accounting for 3,239.4 hours of recorded training activity. A
technical note that details the process for extracting the training
sessions from timestamped measurements is provided in the
supplementary material.  Figure~\ref{study} shows the resulting
observation timeline for each runner and puts the events of the study
on a calendar scale, where triangles denote lab tests, circles denote
field tests and crosses denote training sessions. In what follows, a
training period is defined as the interval between two consecutive
field tests. So, as is also apparent from Figure~\ref{study}, each
runner had 4 training periods.
The figure also shows the number of recorded sessions and number of
training hours within each training period (above and below each
timeline, respectively) and the corresponding totals for the whole
timeline (on the right of each timeline). As seen on
Figure~\ref{study}, some runners have no training records for long
intervals on their timelines. These intervals are either because of
absence from training due to injury or vacation, or due to the runner
failing to deliver their GPS container files to the scientist.

The training speed profiles for each training session shown in
Figure~\ref{study} were calculated from the timestamped GPS
measurements after appropriate imputation of zero speeds. The
respective calculations and the imputation process are detailed in the
technical note provided in the supplementary material.

\section{Modelling running performance}
\label{modelling}

\subsection{An extended power law for running times}


In order to investigate the effects of training and physiological
factors on the running performance, it is assumed that the performance
$Y_{ik}$ (in seconds) at the $i$th field test over distance $D_k$
decomposes as
\begin{equation}
  \label{model}
  Y_{ik} = \tau D_k^{\alpha} e^{\zeta_{i}} e^{\theta_{i}}e^{o_{i}}
  e^{\epsilon_{ik}} \quad (i = 1,
  \ldots, 56; k = 1, \ldots, 3) \,
\end{equation}
where $\tau$ and $\alpha$ are the parameters controlling the power-law
relationship between performance times and distances \citep{katz:99,
  savaglio:2000} and $\epsilon_{ik}$ is an error component with zero
mean. Model~(\ref{model}) extends the established power-law model for
performance, by also taking into account the effect of the runners'
physiological status ($\zeta_i \in \Re$), the effect of training
($\theta_i \in \Re$), and the effects of other factors, for example
psychological, environmental ($o_{i} \in \Re$) that can potentially
influence performance. Model~(\ref{model}) asserts that the mean
performance of the runners decomposes into a distance effect
$D_k^{\alpha}$, a physiological status effect $e^{\zeta_{i}}$, a
training effect $e^{\theta_{i}}$ and the effect $e^{o_{i}}$ of other
unmeasured factors. Note that the inclusion of physiological status
effects also brings runner-specific effects into the model.

In the current study we only have information for the distance, and
the effects of physiological status effect and training. For this
reason and taking into account the careful standardization prior to
all laboratory and field tests (see Subsection~\ref{collection} for
more details), we assume that the effect of other unobserved factors
on the performance across field tests is constant and set
$e^{o_{i}} = 1$.

\begin{table}[t!]
  \begin{center}
    \begin{small}
      \begin{tabular}{lll} \toprule Effect & Parameter & Covariate
        information \\ \midrule Distance ($\alpha \log D_k$) &
        $\alpha$ & Distance (metres) for performance trial \\ \midrule
        Physiological status ($\zeta_{i}$) & $\gamma_1$ & Weight (kg) \\
                                           & $\gamma_2$ & Height (cm) \\
                                           & $\gamma_3$ & Age (years) \\
                                           & $\gamma_4$ & $\dot{V}$O$_{2max}$ (ml$\cdot$min$^{-1}\cdot$kg$^{-1}$) \\
                                           & $\gamma_5$ & $\dot{V}$O$_{2max}$ (km$\cdot$h$^{-1}$) \\
                                           & $\gamma_6$ & Economy (ml$\cdot$kg$^{-1}\cdot$km$^{-1}$) \\
                                           & $\gamma_7$ & Economy (kcal$\cdot$kg$^{-1}\cdot$km$^{-1}$) \\
                                           & $\gamma_8$ & OBLA
                                                          (m$\cdot$s$^{-1}$
                                                          at which
                                                          blood
                                                          lactate
                                                          reaches 4mM)
        \\ \midrule
        Training ($\theta_{i}$) & $\delta_0$  & Average session length (seconds) \\
                                           & $\delta(s)$ & Observed
                                                           training
                                                           distribution
                                                           profile
                                                           (seconds)
        \\ \bottomrule
      \end{tabular}
    \end{small}
  \end{center}
  \caption{The available covariate information that
    is used to characterize each of the effects in model~(\ref{model}).}
  \label{covariates}
\end{table}

\subsection{Definition of physiological status effect}
The physiological status effect in model~(\ref{model}) is assumed to
have the additive form
\begin{equation}
  \label{psysiologyeffect}
  \zeta_{i} = \gamma^Tl_{i}\, ,
\end{equation}
where $\gamma$ is a vector of unknown parameters and $l_{i}$ is the
vector of the laboratory test results at the end of the $i$th training
period. Table~\ref{covariates} lists the available laboratory test
results, which involve measurements on weight, height, age and
physiological status determinants of running performance.

\subsection{Definition of the training effect via training
  distribution profiles}
\label{trainingprofile}
\subsubsection{Training distribution profiles}
\begin{figure}[t!]
  \begin{center}
    \includegraphics[width =
    \textwidth]{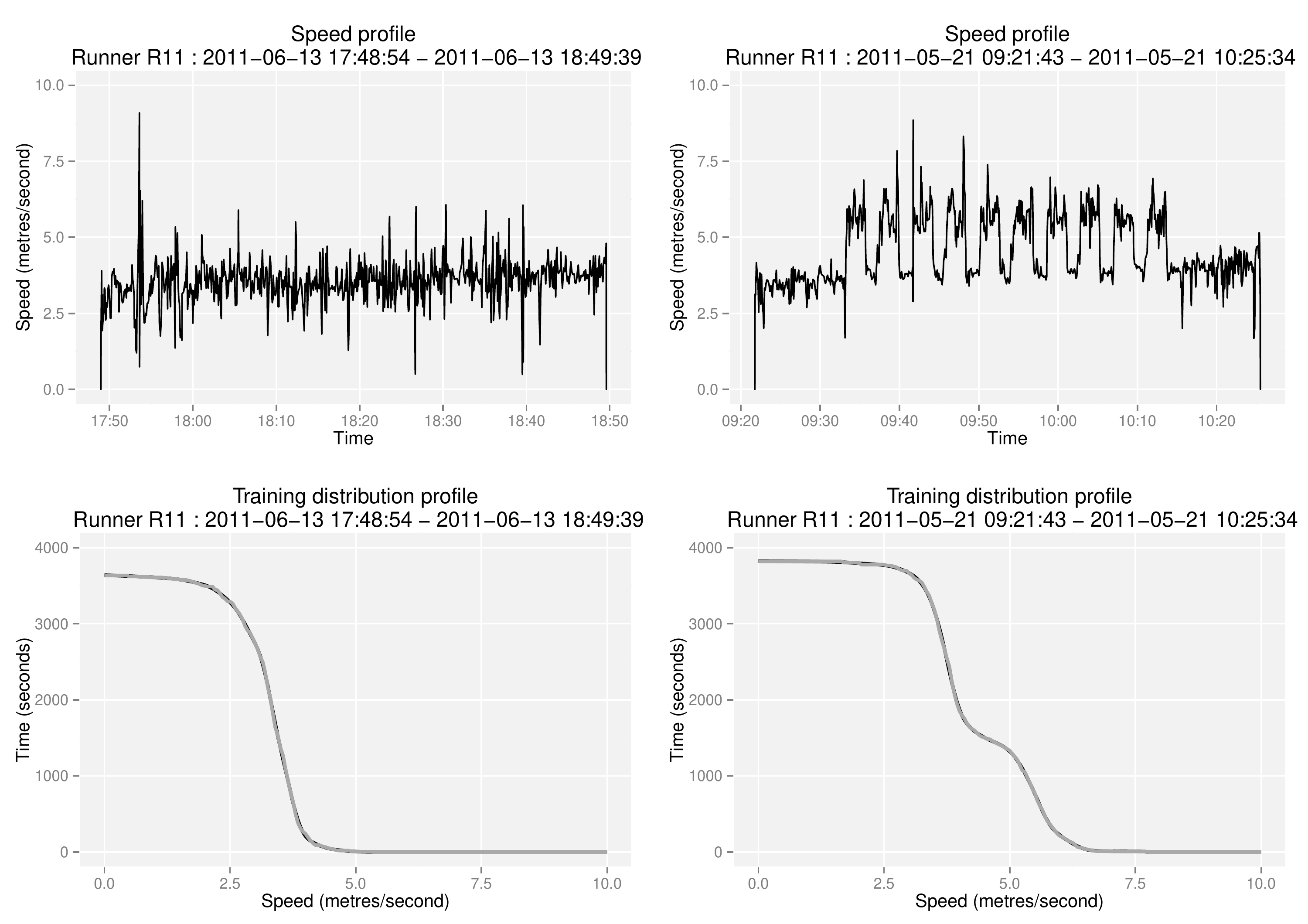}
  \end{center}
  \caption{Two speed profiles (top) and the corresponding training
    distribution profiles (bottom). The black curves in are calculated
    using (\ref{observedTrainingDistribution}) and the grey smoother
    curves result by using a constrained shape additive model
    \citep{pya:14} with Poisson responses.}
  \label{speedprofiles}
\end{figure}

The definition of the training effect in model~(\ref{model}) has to
incorporate an effective summary of the training that took place over
each training period. Using directly some summary of the training
speeds, such as a few quantiles, is an option but the speed profiles
are rather noisy sometimes resulting in extreme speeds (see, for
example, the top row of Figure~\ref{speedprofiles} for two
well-behaved profiles). So, the use of a smoothing procedure is
necessary prior to the calculation of any summaries. This, though,
would require assumptions on the behaviour of runners' speeds during
their regular training sessions as a function of time in order to
determine the right amount of smoothing. Furthermore, any scalar
summary of speeds over the training period does not directly reflect
how the runner planned the allocation of speeds in the sessions within
each training period.

In order to overcome such difficulties in defining the training
effect, we introduce the concept of the training distribution
profile. For a session $u$ that lasted $t_u$ seconds, let $K_v =
\left\{t\, |\, \upsilon_u(t) > v\,, t \in (0, t_u)\right\}$ and
\[
\Pi_{u}(v) = \int_{0}^{t_u}I(\upsilon_u(t) > v)dt = {\rm Length}(K_v)
\, .
\]
We call the curve $\{v, \Pi_u(v)\, |\, v \ge 0\}$, the ``training
distribution profile''. The training distribution profile represents
the training time spent exceeding any particular speed threshold and
is a smoother representation of the allocation of speeds than the
speed profile. Note that $\Pi_u(a) \ge 0$ for any $\alpha \ge 0$, and
that $\Pi_u(a) = t_u$ for any $a < 0$.  In addition, $\Pi_{u}(v)$ is a
necessarily decreasing function of speed.

The observed version of $\Pi_u(v)$ can be calculated using times and
the calculated speeds (see the technical note in the supplementary
material for details on the calculation of speeds) as
\begin{equation}
  \label{observedTrainingDistribution}
  P_u(v) = \sum_{j = 2}^{n_u}(T_{u,j} - T_{u,j-1})I(V_{u,j} > v) \, ,
\end{equation}
where $n_u$ is the number of observations after the imputation process
has taken place, and $V_{u,j}$ is the calculated speed, in meters per
second, at time $T_{u,j}$. Then, for a chosen grid $v_1, \ldots, v_k$
of speed values, $\{v_1, P_u(v_1)\}, \ldots, \{v_k, P_u(v_k)\}$, can
be used to estimate the training distribution profile using smoothing
techniques that respect the positivity and monotonicity of
$\Pi_u(v)$. The top of Figure~\ref{speedprofiles} shows two examples
of speed profiles, one characteristic of an almost constant-pace
training session and one from a high-intensity, interval-based
training session. The corresponding estimated training distribution
profiles are shown in the bottom row of Figure~\ref{speedprofiles}. As
can be seen, the difference in the structure of the two training
sessions alters the shape of the training distribution profiles
correspondingly. The black curves in Figure~\ref{speedprofiles} are
calculated using (\ref{observedTrainingDistribution}) and the grey,
smoother curves result from fitting a shape constrained additive model
with Poisson responses \citep{pya:14} to ensure that the relationship
between the smoothed version of $P_u(v)$ and $v$ is still monotone
decreasing. The fitting is done using the \texttt{scam} package in
{\bf R} \citep{rcore:15,scam}. Note that the resulting smoothed
version is almost indistinguishable from $P_u(v)$ and from visual
inspection (not reported here) this is the case for all recorded
sessions in the data. For this reason, all subsequent analysis uses
directly the smoothed versions.

Any training sessions that correspond to estimated training
distribution profiles with more than 125 seconds above 8
m$\cdot$s$^{-1}$ were dropped from the analysis as errors in data
collection, because they exceed the world record speed for 800 metres
(7.93 m$\cdot$s$^{-1}$ for 800m, David Rudisha, London Olympic Games,
9 August 2012). There were $18$ such sessions in the data and they
have all been identified with the participants as bicycle rides or
instances where the participants did not switch off their GPS device
before driving their car or riding their bicycle after the end of
training session.

\subsubsection{Training effect}
For defining the training effect in model~\ref{model}, we assume that
the average structure of the available training sessions within a
training period (see Figure~\ref{study}) approximates well the average
training behaviour of runners for all the training sessions that took
place in that period. We, further, assume that there is an unknown
real-valued weight function $\delta(s)$, that weights the time spent
at each speed according to its importance in determining performance.
\begin{equation}
  \label{trainingeffect}
  \theta_{i} = \delta_0 \bar{t}_{i} - \int_{0}^{\infty}
  \bar{P}'_{i}(s)\delta(s)ds \quad (i = 1, \ldots, 56)\, ,
\end{equation}
where $\delta_0 \in \Re$ is an unknown parameter and $\bar{P}'_{i}(s)
= d\bar{P}_{i}(s)/ds$ with $\bar{P}_{i}(s)$ being the average of the
smoothed training distribution profiles for the $i$th training
period. The quantity $\bar{t}_{i} = \bar{P}_{i}(-1) = \int_0^\infty
\bar{P}'_{i}(s) ds$ is the average session length in the $i$th
period.

\begin{figure}[t!]
  \begin{center}
    \includegraphics[width = \textwidth]{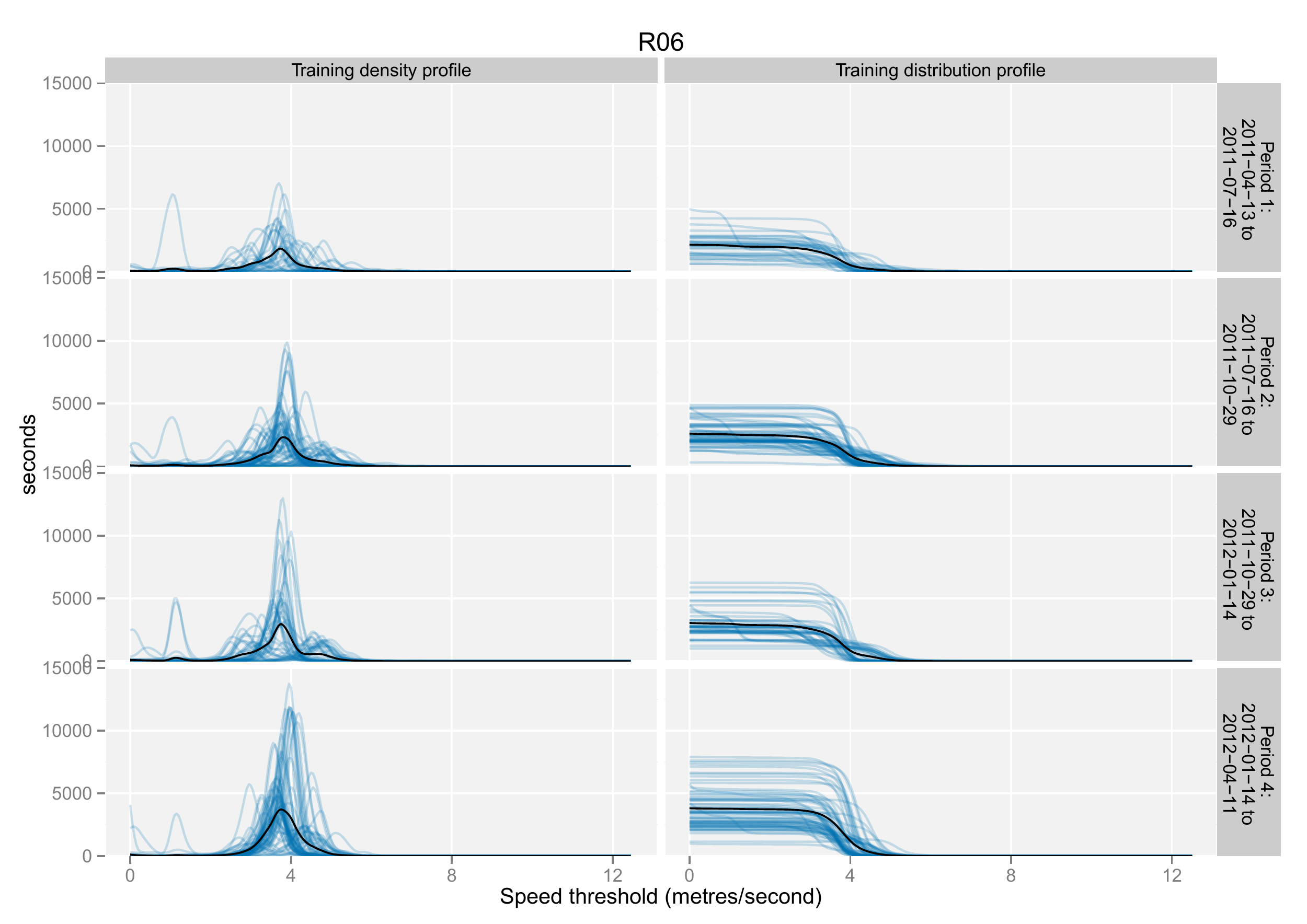}
  \end{center}
  \caption{The observed, smoothed training distribution profiles for
    all sessions in the training periods for runner R06 (right) and
    their negative derivatives (left). The average curves are shown in
    black.}
  \label{trainingdistribution}
\end{figure}

As a final note, in contrast to individual session speed profiles the
concept of training distribution profiles appears well-suited for
visualising large volumes of training session data. For example, the
right panel of Figure~\ref{trainingdistribution} shows the estimated
training distribution profiles for all sessions in the training
periods of runner R06. The left panel shows the negative derivatives
of those and reveals the clear concentration of training at around 4
m$\cdot$s$^{-1}$ which would have otherwise been hard to
visualise. The average curves are shown in black.

\section{Estimation}
\label{estimation}
\subsection{Multi-resolution elastic net}
\label{multi}
Table~\ref{covariates} lists the available covariate information that
is used to characterize each of the effects in
model~(\ref{model}). Taking logarithms on both sides,
model~(\ref{model}) is a linear regression model with response the
logarithm of the performance at the $k$th distance, $10$ scalar
parameters (one for each scalar covariate in Table~\ref{covariates}
and the intercept $\log\tau$) and a functional parameter
$\delta(s)$.

We wish to be able to use that regression model to simultaneously
assess the importance of scalar and functional covariates, also taking
into account the correlation between the scalar covariates, and that
between the scalar and the functional covariates. Particularly, the
importance of the functional covariates can be assessed by identifying
the training speed intervals that are important for performance
changes. If there were no scalar covariates, one way to do so is the
FLiRTI method in \citet{james:09}, which induces sparseness
simultaneously on $\delta(s)$ and on derivatives of it of a preset
order.


In order to simultaneously identify important training speeds and
important training covariates, we propose an alternative procedure
which we term the multi-resolution elastic net and which consists of
the following steps. For each resolution $G$ from a set of
resolutions:
\begin{itemize}
\item[i)] Partition the union of the observed domains of the
  functional covariate across observations into $G$ intervals of the
  same length.
\item[ii)] Construct $G$ covariates by calculating a summary of the
  functional covariate for each observation and on each interval
  (e.g. integral, difference at endpoints and so on).
\item[iii)] Apply the elastic net \citep{zou:05} on the covariates
  constructed in step b) along with any other available scalar
  covariates.
\end{itemize}
The $L_2$ penalty of the elastic net in step iii) controls for the
extreme collinearity that the covariates of step ii) and any other
scalar covariates can have, and the $L_1$ penalty of the elastic net
imposes sparseness, if necessary. Note that the above makes no direct
assumption on the ordering of the regression parameters corresponding
to the functional covariate, as a functional regression approach would
do. Instead, multi-resolution elastic net takes advantage of the
grouping properties of the elastic net \citep[][Section~2.3]{zou:05}
for the formation of contiguous groups of non-zero estimates for the
parameters of the highly correlated summaries of the functional
covariates in step ii). In this way, if the non-zero elastic net
coefficients of those parameters form contiguous groups, then there is
strong evidence that the corresponding intervals are important for the
response. A further persistence of those contiguous groups as the
resolution $G$ increases will further strengthen any conclusions.

For the current application, in steps i) and ii) of the
multi-resolution elastic net we replace (\ref{trainingeffect}) with
the discretized version
\begin{equation}
  \label{trainingeffect1}
  \theta_{i} = \delta_1 \bar{t}_{i} + \sum_{g = 1}^{G}
  \left\{\bar{P}_{i}(v_{g-1}) - \bar{P}_{i}(v_{g})\right\} \delta_{g} \quad (i = 1, \ldots, 56)\, ,
\end{equation}
where $\bar{P}_{i}(v_{g-1}) - \bar{P}_{i}(v_{g})$ is the average
training time spent in the $g$th speed interval before the $i$th
end-of-period field test, and $\{v_0, \ldots, v_G\}$ is an equi-spaced
grid of speeds with $v_0 = 0$ and $v_G = 12.5$ for some resolution
$G = 1, 2, \ldots$.  Then specification (\ref{trainingeffect1}) and
the scalar covariates are handled simultaneously in step iii) of the
procedure.  This last step will return estimates on
$\beta = (\log\tau, \alpha, \gamma_1, \ldots, \gamma_8, \delta_0,
\delta_1, \ldots, \delta_G)^T$
and, hence, estimate the mean of the logarithmic performance
\begin{equation}
  \label{logmu}
  \log{\mu_{ik}(\beta; G)} = \log\tau + \alpha\log D_k  + \gamma^T l_i +
  \delta_0 \bar{t}_i + \sum_{g = 1}^{G}
  \left\{\bar{P}_{i}(v_{g-1}) - \bar{P}_{i}(v_{g})\right\} \delta_{g} \, ,
\end{equation}
for various resolutions $G$.

\subsection{Selection of tuning constants and of optimal resolution}
The above setup requires the determination of an optimal resolution
$G$ for the training effects and the selection of the two tuning
constants of the elastic net. In order to do so, the data set is
first split into an estimation and a test set (the commonly used
terminology for the ``estimation set'' is ``training set'' but we
diverge from that in order avoid a terminology clash with the training
effect $\theta_i$). Then, for each resolution, the tuning constants of
the elastic net are selected as the ones that minimise the mean
squared prediction error estimated using 10-fold cross-validation
repeated for 10 randomly selected fold allocations in the estimation
set. The selection of the two tuning constants of elastic net via
cross-validation was implemented using the \texttt{caret}
\citep{kuhn:08} and \texttt{elasticnet} \citep{elasticnet:12} {\bf R}
packages. The resolution $G$ is then determined as the one that
minimises the mean squared prediction error estimated using the test
set. Another outcome of this process is the estimated model for the
chosen optimal resolution, which can be used for predictive purposes.

\subsection{Estimation and test set}
The six records that correspond to the last period for runners R12 and
R14 were dropped from the data as uninformative because that period
contained no or only one training records (see
Figure~\ref{study}). The resultant data set of 162 observations was
then split into an estimation and test set. The test set is built from
the records of 4 randomly selected
runners. 
The reason for selecting amongst runners instead directly amongst
records is to avoiding choosing an overoptimistic model in terms of
prediction (note that the physiological status covariate information
is repeated across distances in model~(\ref{model})). The 114 records
for the remaining 10 runners form the estimation set.

\section{Results}
\label{results}

\begin{figure}[t!]
  \begin{center}
    \includegraphics[width = \textwidth]{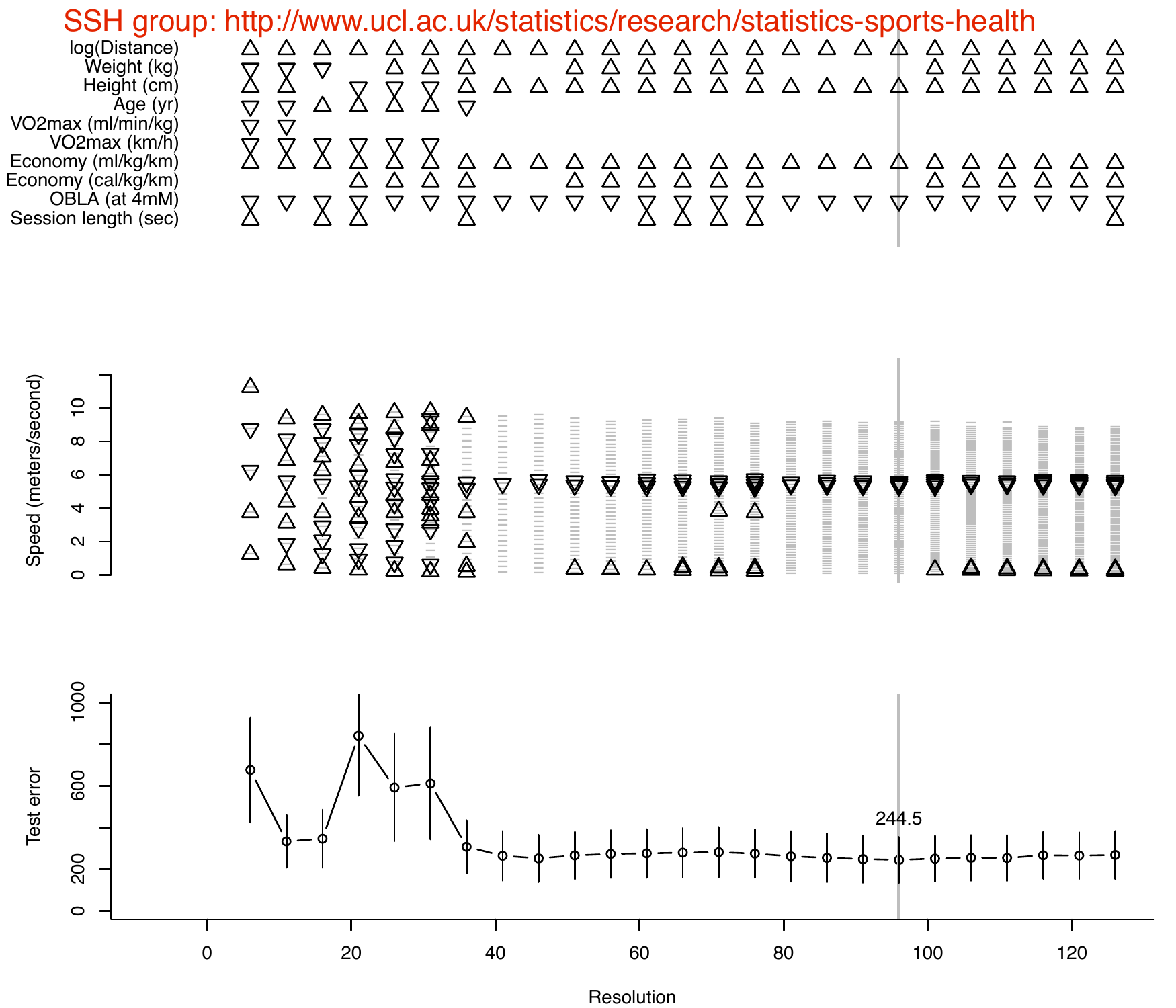}
  \end{center}
  \caption{The non-zero estimated parameters for the chosen tuning
    constants of the elastic net for resolutions
    $G \in \{5, 10, \ldots, 125\}$. The vertical segments represent
    approximate normal-theory $95\%$ confidence intervals.}
  \label{significance}
\end{figure}

\subsection{Distance and physiological status effects}
Figure~\ref{significance} shows the non-zero estimated parameters for
the chosen tuning constants of the elastic net for resolutions
$G\in\{5, 10, \ldots, 125\}$, where the maximal resolution of $125$
corresponds to speed intervals of length $0.1$ m$\cdot$s$^{-1}$
each. The figure provides a quick assessment of the relevance of
each covariate. The symbols $\bigtriangledown$ and $\bigtriangleup$ in
Figure~\ref{significance} indicate negative and positive elastic net
estimates respectively. The signs of the estimated parameters are all
as expected.

As is apparent from the top plot in Figure~\ref{significance} the
distance of the field test, and the physiological status covariates of
height (cm), running economy (ml$\cdot$kg$^{-1}\cdot$km$^{-1}$) and
running speed at OBLA are influential determinants of performance,
irrespective of the resolution used. Particularly, the model indicates
that without varying the training effects, shorter runners with higher
speeds at OBLA and superior running economy
(ml$\cdot$kg$^{-1}\cdot$km$^{-1}$) perform better over a fixed race
distance. The importance of each of OBLA, economy and height for
running performance has been established and is consistent with
previous study findings.  For example, marathon performance has been
shown to be predicted by running speed at OBLA and \citet{sjodin:81}
suggest that OBLA is reflective of the underlying physiological status
of endurance runners. The influence of running economy on endurance
performance has been previous studied in \citet{conley:80} for a
cohort of highly-trained runners similar to those of the present
study. In striking similarity to the current analysis,
\citet{conley:80} also found significant evidence that runners'
economy associates with performance whereas $\dot{V} O_{2max}$ is
not. The relevance of height for endurance performance does not appear
to have been widely studied. In this respect, \citet{bale:86} examined
a number of characteristics in a group of elite runners by dividing
them according to by their best 10 kilometres performance time. They
found significant evidence that the runners with larger performance
times (slower) than the group's median were taller than those who had
run faster.

\subsection{Training effects}
The middle plot in Figure~\ref{significance} indicates that the time
spent training at speeds in the approximate interval from 5.3 to 5.7
m$\cdot$s$^{-1}$ is influential to the improvement in performance. The
plot also shows that this result persists across all resolutions
considered by the multi-resolution elastic net, and its importance is
enhanced by the fact that within that interval and for all resolutions
the non-zero estimates form a contiguous group. To the authors'
knowledge, this finding is the most specific to date in terms of
analysing the contribution of training to subsequent performance.

Overall, the current analysis allows us to identify influential
training speeds for subsequent performances within any training
programme.  This is in contrast to previous research where, typically,
training speeds have been defined according to an underlying
physiological model adopted prior to the commencement of the study,
e.g. as percentages of the speed at $\dot{V}O_{2max}$ or OBLA. For
example, \citet{galbraith:14} identify 4.3 m$\cdot$s$^{-1}$ as the
speed at OBLA, and then examine whether training at all speeds higher
or lower than this links to changes in performance.


\subsection{A predictive equation for performance}
The records in the test set were used to calculate the squared
prediction test error
\[
\sum_{i \in \mathcal{T}} \sum_{k = 1}^3 \left(y_{ik} -
  \mu_{ik}(\hat\beta; G) \right)^2
\]
for each $G \in \{5, 10, \ldots, 125\}$, where $\mu_{ik}(\beta; G)$ is
the exponential of (\ref{logmu}) and $\mathcal{T}$ is the subset of
$\{1, \ldots, 162\}$ that contains the indices of the $48$
observations in the test set. The elastic net estimates $\hat\beta$
have been rescaled as in \citet{zou:05} to avoid over-shrinkage due to
the double penalization that takes place in elastic net. The test
errors and the corresponding 95\% normal theory intervals ($\pm$ 1.96
standard deviations) are shown at the bottom plot in
Figure~\ref{significance}. The minimum test error is achieved for
$G = 95$.

\begin{figure}[t!]
  \begin{center}
    \includegraphics[width = \textwidth]{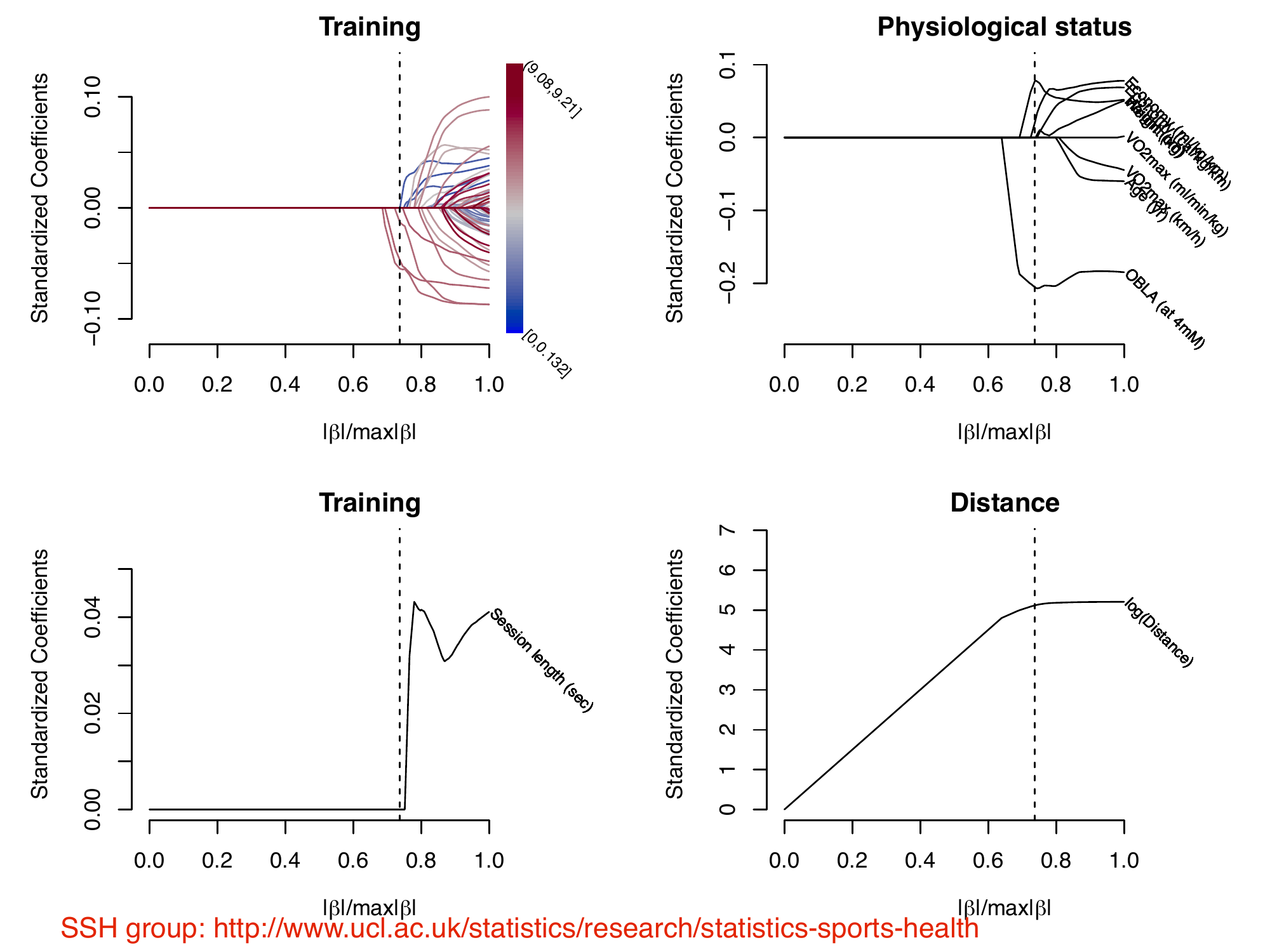}
  \end{center}
  \caption{The elastic net solution paths for resolution $G = 95$
    versus the fraction of the $L_1$ norm. The fit identified by the
    dashed line on the solution paths corresponds to the non-zero
    estimates at $G = 95$ in Figure~\ref{significance}.}
  \label{paths}
\end{figure}

Figure~\ref{paths} shows the elastic net solution paths for that
resolution as a function of the fraction of the $L_1$
norm. Particularly, the fit identified by the dashed line on the
solution paths corresponds to the non-zero estimates at $G = 95$ in
Figure~\ref{significance} (grey line). The elastic net estimates for
the $G = 95$ fit can be used to form the expression that predicts
performance (in seconds) using the race distance, physiological status
determinants as measured in the laboratory, and the average training
distribution profile. This expression has been estimated to be
\begin{align}
  \label{predict}
  & 0.1310 \text{``Distance (m)"}^{1.0568} \\ \notag
  & \exp\left\{0.1007\text{``Height (m)"} + 0.1657\text{``Economy
    (L$\cdot$kg$^{-1}\cdot$km$^{-1}$)"} -0.0159\text{``OBLA
    (km$\cdot$h$^{-1}$)"}\right\} \\ \notag
  & \exp\left\{-0.0078t_1 -0.0279t_2 -0.0307t_3\right\} \, ,
\end{align}
where $t_1, t_2, t_3$ are the training period average times in minutes
spent training within the speed intervals $(5.26,5.39]$,
$(5.39,5.53]$, $(5.53,5.66]$, respectively. In order to reduce
rounding error the units adopted in the above equation are rescaled
from those in Table~\ref{covariates} so that height is calculated in
metres and economy in litres per kilogram per kilometre. The exponent
of Distance in (\ref{predict}) is in agreement with previous studies
\citep{katz:99, savaglio:2000} where it has been found to be around
1.1. Expression~(\ref{predict}) can be used to determine the
performance of an endurance runner for a specified race distance, by
supplying the runner's height, the measurement of economy and OBLA
from a laboratory test prior to the race, and the average time spend
training at the specified speed intervals during the period prior to
the race.

\section{Discussion}
\label{discussion}
A multiplicative effects model has been used to link the training and
physiological status of highly-trained endurance runners to their best
performances. The model extends previous work that uses the power-law
to describe the relationship between performance times and distance,
by also including information on the physiological status of the
runner as measured under laboratory conditions, and the runners'
training as extracted directly from GPS timestamped distance records
for the period prior to the performance assessment.

The relevance of the training and physiology covariates in the
model was assessed using multi-resolution elastic net, which is
described in detail in Subsection~\ref{multi}. We argue that
multi-resolution elastic net is a useful procedure to quickly check
for the existence of influential intervals on the domain of one or
more functional covariates in the presence of other scalar
covariates. To provide some more evidence for our claim, a
reproducible analysis of a popular dataset in functional regression is
revisited in the supplementary material and the results of
multi-resolution elastic net are sensible and in agreement with those
of the FLiRTI procedure of \citet{james:09}. Note that, the latter
procedure is designed to handle regression models with only a single
functional covariate. This is in contrast to multi-resolution elastic
net that can simultaneously handle arbitrary number of scalar and
functional covariates.

An important and novel aspect of the present study was to determine
the direct effect of training on performance. Multi-resolution elastic
net identified that the time spent training between 5.3 and 5.7
m$\cdot$s$^{-1}$ relates to improvements in the performance of the
endurance runners. Another important aspect of the current study is
that it was able to reproduce well-established relationships between
performance and other physiological status measures
\citep[][]{conley:80, sjodin:81, bale:86}, without limitations posed
by an underpinning physiological or training model. Specifically, it
was found that without varying training effects, shorter runners with
higher speeds at OBLA and superior running economy
(ml$\cdot$kg$^{-1}\cdot$km$^{-1}$) are found to perform better.

As mentioned in the Introduction section, the effect of training on
performance has been modelled in studies like \citet{banister:75} and
\citet{busso:03} using a training impulse response model. The model
adopted in those and similar studies, though, cannot be used for
predictions beyond the training session and runner
studied. Furthermore, the training inputs for the models in those
studies are aggregated into arbitrary units thereby limiting their
value to theoretical rather than practical applications. Indeed
subsequently, \citet{busso:06} stress the need for the development of
new modelling strategies by stating that {\em ``it is likely that the
  expected accuracy between model prediction and actual data will
  greatly suffer from the simplifications made to aggregate total
  training strain in a single variable and, more generally, from the
  abstraction of complex physiological processes into a very small
  number of entities''}. In this respect, expression~(\ref{predict})
of our work generalises beyond the study design described in
Section~\ref{studydata} and, in addition, it has been chosen as the
best in terms of the predictive quality from the models arising from
different levels of aggregation of the training inputs.

Training programmes and the individual sessions within them can be
complex. This complexity can make it difficult to visualise and
analyse a large training dataset without prior transformation or
simplification. The current work contributes in this direction with
the introduction of the concept of the training distribution
profile. The training distribution profile promotes clear and
straightforward visualisation of large volumes of training data (see
for example Figure~\ref{trainingdistribution}). More importantly, the
training distribution profile allows the use of a wide range of
contemporary statistical methods for the modelling of training
data. For example, the training distribution profiles or their
derivatives can directly be used as responses or covariates in
functional regression models \citep[see, for example][Chapters 15--17
for details]{ramsay:05} and/or for the detection of training regimes
and changes in training practices, for example by a cluster analysis
\citep[]{james:03}. Furthermore, a more fruitful, if not more
involved, analysis would, for example, take into account the
variability of the training distribution profiles and/or their
derivatives, as well as any serial correlation between them. The
methods in \citet{bathia:10} seem to provide a good basis in this
direction. Given the widespread use of wearable data recording devices
in training, distribution profiles of other aspects of a runner's
training can also be produced, such as heart rate distribution
profiles and/or power-output profiles in cycling for example. Their
influence on performance can then be determined using similar
procedures as in the present study.

Overall, we try to reverse the prevailing scientific paradigm for
investigating the effects of training on performance. Rather than
evaluating the effects of a pre-specified training programme, we
instead identify those aspects of training that link to a measurable
effect. This presents an exciting and promising new approach to
developing training theory. Importantly, the ability to identify
important speeds presents an obvious focus for subsequent training
interventions on the performance of endurance runners, and motivates
further subject-specific work on the design and the study of the
effectiveness of such interventions. If this work is successful, it
has the potential to lead to the development of a new model of
training which could be tuned towards maximising performance gains, or
enhancing the health benefits arising from a prescribed amount of
exercise.

\section{Supplementary material}
The supplementary material is appended at the end of the current
preprint, and contains a reproducible analysis of the Canadian
weather data (see, for example, \citealt[Section~6]{james:09} or
\citealt[Section~1.3]{ramsay:05}) using multi-resolution elastic net
and the FLiRTI procedure of \citet{james:09}. The supplementary
material also contains a technical note that details the process for
extracting the training sessions and speed profiles from timestamped
GPS measurements.

R scripts that reproduce the analyses undertaken in this manuscript
are available upon request to the authors.

\bibliographystyle{chicago}



\section*{Supplementary material}

Technical note on the extraction of speed profiles from raw GPS data

Ioannis Kosmidis
Department of Statistical Science, University College London
Gower Street, London, WC1E 6BT, United Kingdom
`i.kosmidis@ucl.ac.uk`

and

Louis Passfield
Endurance Research Group,
School of Sport and Exercise Sciences, University of Kent,
Chatham Maritime, Chatham, Kent, ME4 4AG, United Kingdom
`l.passfield@kent.ac.uk`

June 3, 2015

# 1 GPS container files

Whenever the GPS device was reset, a new XML container file (extension .tcx) was created which included all the observations since the previous reset. There were several cases where the GPS devices were not reset by the runners which resulted in several days of training being collected in the same file. Each GPS container file contains timestamped measurements of cumulative distance calculated using latitude and longitude information. The sampling rate between resets was variable, determined by a proprietary algorithm; quoting from the GPS device manual *"The Forerunner uses smart recording. It records key points where you change direction, speed, or heart rate"* (Garmin Ltd., 2013).

# 2 Identification of training sessions

The timestamps and cumulative distances for the complete observation period and for all runners were extracted from the available GPS container files using the XML package (Lang, 2013) in **R** (R Core Team, 2015). The timestamps contain information on the year, month, day, hour, minute, and second that each observation is taken in Greenwich Mean Time. The training distances were recorded in metres. The extracted dataset comprised of 2,499,894 informative observations in the sense that each had a complete timestamp and distance record. The ordered set of timestamps within the observation period of each runner was then used to group the data into training sessions. Particularly, any two consecutive timestamps that were more than 2 hours apart were considered to be the last and the first observations of two consecutive sessions. In

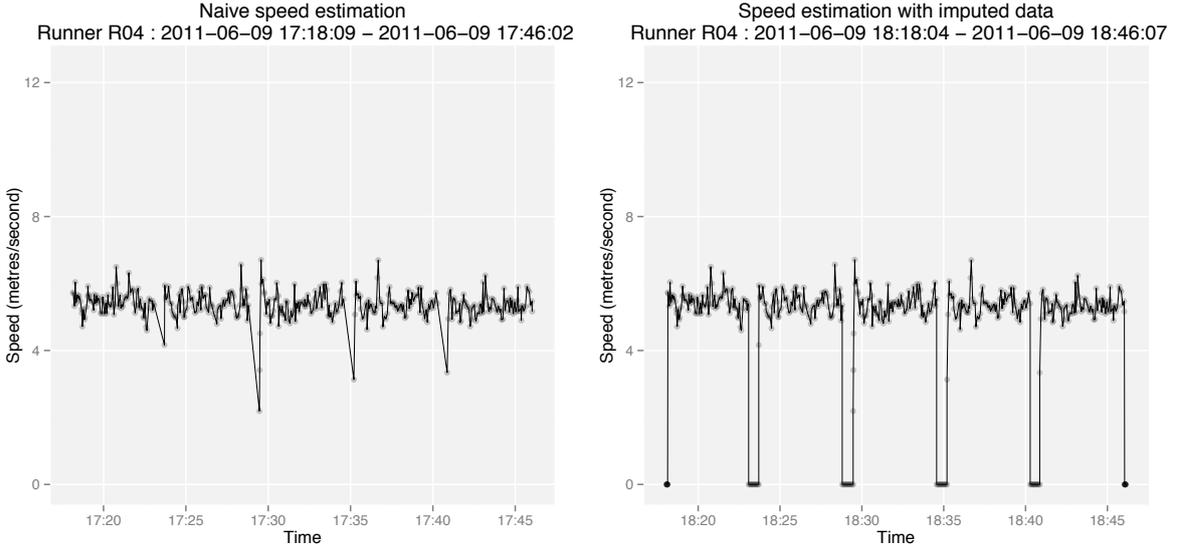

Figure 1: The speed profile of a session in the data as calculated ignoring the gaps in the session distance-time data (left) and after the imputation process in Figure 2 for $l_{(\text{gap})} = 30$ seconds, $l_{(\text{skip})} = 5$ seconds and $m = 10$.

this way, the data were grouped into 3,525 distinct training sessions. Then, the scatterplots of cumulative distance versus time for each of the identified sessions were visually inspected before identifying 56 sessions as being the result of accidental or inappropriate use of the GPS device. These sessions were removed from the data. The remaining 3,469 sessions account for 3,239.4 hours of recorded training activity.

## 3  Imputation of gaps in session data

In several instances, there were consecutive observations within sessions with timestamps that were several minutes apart. This can happen either because the runner stops the recording during the training session, or because the proprietary algorithm that determines the sampling rate detects that there is no change in latitude and longitude (i.e. the device is not moving), and sampling rate is reduced considerably. A naive approach where the speeds are calculated from the available distance-time data may therefore wrongly lead to calculated positive speeds at times where the runner was not moving. The left panel of Figure 1 illustrates this for a session in the data, where the large gaps between observations need to be taken into account when estimating speeds.

Suppose that there was a preset minimum sampling rate that the GPS device used when latitude and longitude did not change and let $l_{(gap)}$ be the maximum length of time between two observations that corresponds to this minimum sampling rate. The working assumption we make is that if two consecutive time points $T_{u,j}$ and $T_{u,j+1}$, in a session $u$ are more than $l_{(\text{gap})}$ seconds apart then the runner has stopped moving for an interval between those time points. Under this assumption, a simple imputation process can be developed that takes care of large gaps in the session data by imputing $m$ observations between distant time points. For any session $u \in \{1, \ldots, 3469\}$ and for any consecutive time points with $T_{u,j+1} - T_{u,j} > l_{(gap)}$ and for $l_{(\text{skip})}$ such that $0 < l_{(\text{skip})} < l_{(\text{gap})}$,

1. define $T^*_{u,j} = T_{u,j} + l_{(\text{skip})}$
2. define $h = (T_{u,j+1} - T_{u,j} - 2l_{(\text{skip})})/m$

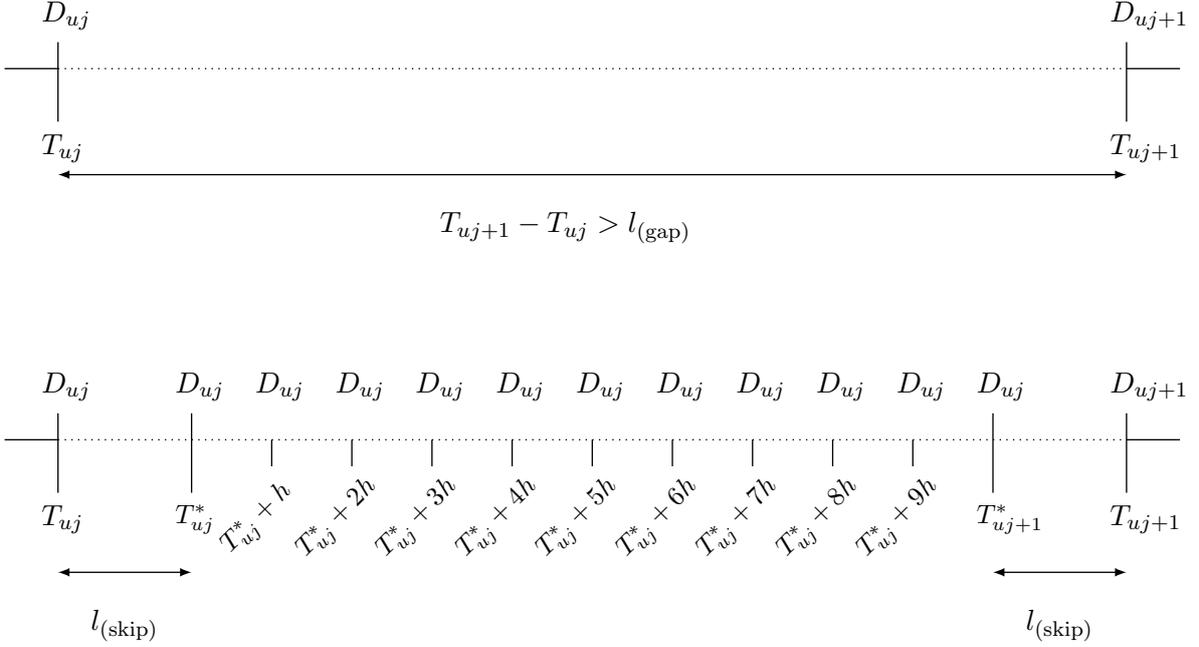

Figure 2: A pictorial representation of the imputation process for $m = 10$.

3. for $j \in \{0, 1, \ldots, m\}$, impute the observations $(T^*_{u,j} + kh, D_{u,j})$ ($k = 0, 1, \ldots, m$) in the session data, where $D_{u,j}$ is the recorded cumulative distance at time $T_{u,j}$.

Figure 2 provides a pictorial representation of the above steps.

The details of the proprietary algorithm that the device is using to determine the sampling rate are not publicly available. Hence, it is not possible to know the value of $l_{(gap)}$ or even if the algorithm makes use of any such threshold. Nevertheless, a good value of $l_{(gap)}$ for the imputation process can be determined using data from 3 of the 56 sessions that have been discarded earlier. Specifically, there are three instances when runner R11 activated the GPS device whilst sleeping and hence all recorded cumulative distances are zero. The only data that the proprietary algorithm of the GPS device had available to use in those instances were the timestamps and the heart rate. Figure 3 shows the heart rate at $T_{u,j}$ versus $T_{u,j+1} - T_{u,j}$ (in log-scale) in each of those instances. A two-dimensional kernel density estimate is overlaid for reference. Note that for recorded heart rates above 60 bits per minute (bpm) the GPS device rarely goes beyond 30 seconds between consecutive observations, when not moving. Given that the average heart rate of the runner in his recorded training sessions is around 140 bpm and that the GPS device must increase its sampling rate for non-zero speeds, $l_{(gap)} = 30$ seconds seems a reasonable assumption. Furthermore, as the runners rarely stop running abruptly or start running immediately we set $l_{(skip)} = 5$ seconds.

## 4 Speed profiles

The imputation process of the previous section is used with $m = 10$. If the speed at time $t$ is $v_u(t) = d\Delta(t)/dt$ (in m·sec$^{-1}$), with $\Delta(t)$ the distance covered at time $t$, then the speed at time $T_{u,j}$ can be approximated using first-order finite differences as

$$V_{u,j} = \frac{D_{u,j} - D_{u,j-1}}{T_{u,j} - T_{u,j-1}} \quad (j = 2, \ldots, n^*_u; u = 1, \ldots, 3469), \tag{1}$$

where $n^*_u$ is the number of observations in the session, after the imputation process took place. After all speeds have been estimated, zero speeds are imputed at $m = 10$ equidistant time

3

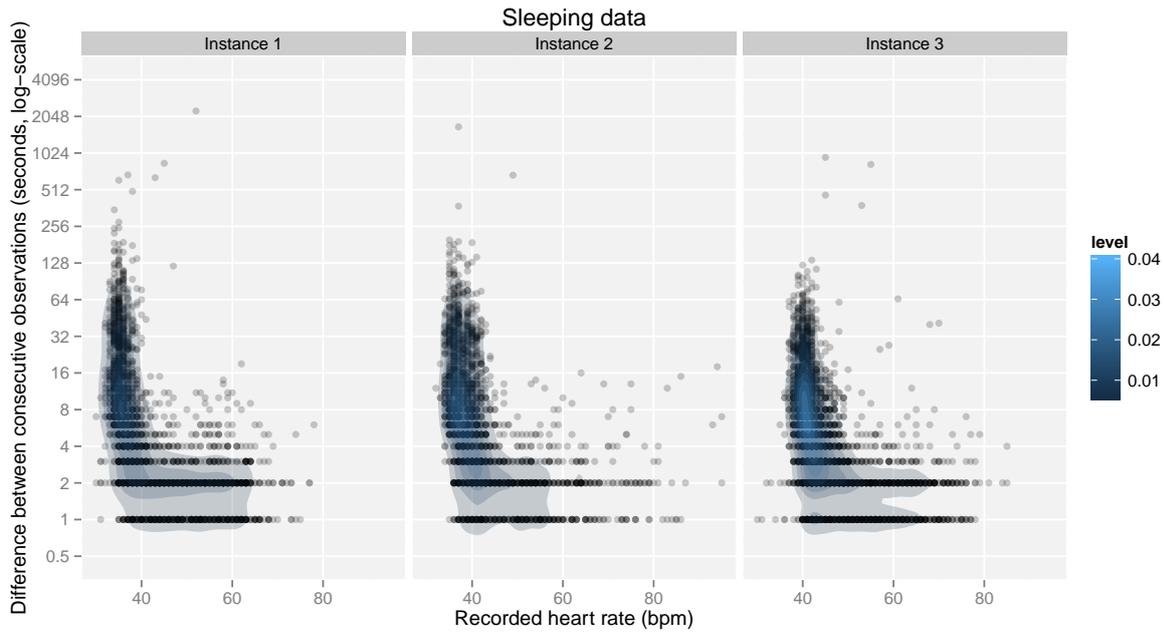

Figure 3: Sleeping data used for estimating $l_{(\text{gap})}$. Heart rate (in bbm) at $T_{u,j}$ versus $T_{u,j+1} - T_{u,j}$ (in minutes, log-scale) for each of the three instances where a runner had an active GPS device whilst sleeping. A two-dimensional kernel density estimate is overlaid for reference.

points 5 seconds before the beginning of the session and at $m = 10$ points at 5 seconds after its end. This results in $n_u = n_u^* + 20$ observations. The right plot in Figure 1 shows the resulting estimated speed profile.

Supplementary material for

Linking the performance of endurance runners to training and physiology effects via multi-resolution elastic net

# Multi-resolution elastic net and FLiRTI for the Canadian weather data

Ioannis Kosmidis
Department of Statistical Science, University College London
Gower Street, London, WC1E 6BT, United Kingdom
i.kosmidis@ucl.ac.uk

and

Louis Passfield
Endurance Research Group,
School of Sport and Exercise Sciences, University of Kent,
Chatham Maritime, Chatham, Kent, ME4 4AG, United Kingdom
l.passfield@kent.ac.uk

June 3, 2015

## 1 R session

The current report assumes that the **R** (R Core Team, 2015) packages `lpSolve`, `fda`, `doMC`, `elasticnet`, `ggplot2` and `plyr` are already installed to an active library. The following code chunk will load those packages into the R session and also register 4 cores for performing some of the computation in parallel.

```r
library(lpSolve)
library(fda)
library(doMC)
library(elasticnet)
library(ggplot2)
library(plyr)
registerDoMC(cores = 4)
```

## 2 Canadian weather data

The analysis of the Canadian weather data (see, for example, James et al. 2009, Section 6 or Ramsay and Silverman 2005, Section 1.3) is revisited here in order to illustrate the process of multi-resolution elastic net in a simple setting with one functional covariate. We also compare the results to those from the FLiRTI (functional linear regression that's interpretable) approach of James et al. (2009). The method of James et al. (2009) has been developed for regressions with a single functional covariate (and no scalar covariates), and, as multi-resolution elastic net, results

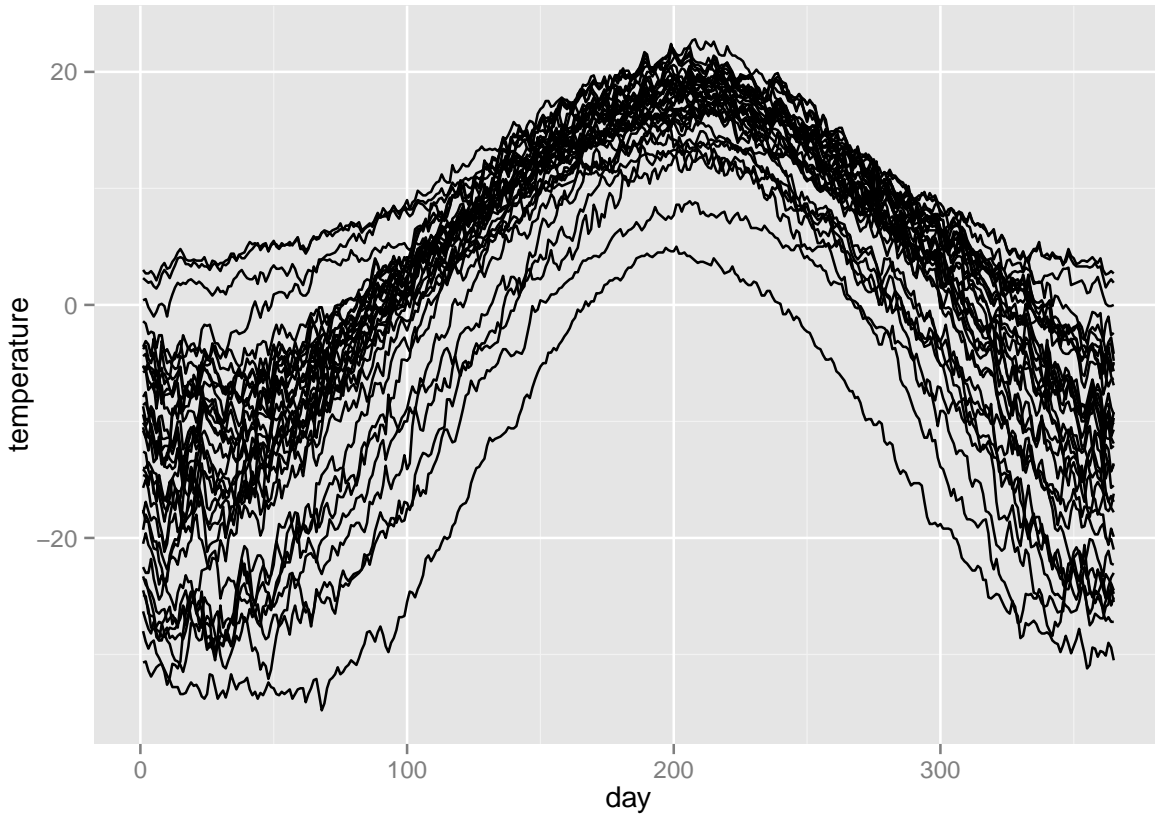

Figure 1: Daily temperature profiles for the 35 weather stations.

in interpretable estimates of the functional parameter by imposing sparseness simultaneously on it and on preset-order derivatives of it.

The Canadian weather data is provided with the `fda` R package (Ramsay et al., 2014) and consists of one year of daily temperature measurements from each of 35 Canadian weather stations. The total annual precipitation at each weather station is also provided. Denote the temperature at day $m$ as $T_i(m)$ and the total annual precipitation as $P_i$ ($i = 1, \ldots, 35$). The values of $T_i(m)$ are available for $m \in \{1, \ldots, 365\}$. Figure 1 shows the daily temperature profiles for the 35 weather stations.

```r
temperatures <- as.data.frame(CanadianWeather$dailyAv[, , "Temperature.C"])
Xdf <- stack(temperatures)
names(Xdf) <- c("temperature", "station")
Xdf$day <- seq.int(nrow(temperatures))
ggplot(Xdf) + geom_line(aes(x = day, y = temperature, group = station))
```

James et al. (2009, Section 6) illustrate the effectiveness of the FLiRTI approach by fitting the functional linear regression model

$$logP_i = \beta_0 + \int_0^{365} \beta(m) T_i(m) dm + \epsilon_i \quad (i = 1, \ldots, 35), \tag{1}$$

where $\beta_0$ is a scalar parameter, $\beta(m)$ is a functional parameter and $\epsilon_i$ is an error with zero mean.

## 2.1  FLiRTI results

In order to estimate model (1) via the FLiRTI approach we used the R code that the authors of James et al. (2009) provide (see http://www-bcf.usc.edu/~gareth/research/flrti and

`http://www-bcf.usc.edu/~gareth/research/flrtidoc.pdf` for more details; all URLs provided above were correct as of June 3, 2015). For the application of FLiRTI we are required to make choices on the basis for $\beta(m)$ and on three tuning constants. We choose a simple grid basis of dimension equal to the number of points that the value of the functional covariate is recorded at (see, for example James et al., 2009, Section 2, for details) and, as in James et al. (2009, Section 6) we choose to restrict the $\beta(m)$ and the third derivative of it, in order to get a sparse, piece-wise quadratic estimate.

```r
# Source the FLiRTI functions
source("http://www-bcf.usc.edu/~gareth/research/flrti")
# Define the response and the functional covariates
Y <- log(colSums(CanadianWeather$dailyAv[,,"Precipitation.mm"]))
Xf <- t(temperatures)
monthLetters <- c("J", "F", "M", "A", "M", "J", "J", "A", "S", "O", "N", "D")
# Settings for the tuning constants
settingsF <- expand.grid(sigma = seq(2, 34, 8)/1000,
                         weight = seq(1, 91, 30)/100,
                         deriv = 3)
# Fits using FLiRTI
modelfits <- alply(settingsF, 1, function(setting) {
   flrti(Y = Y, X = Xf,
         sigma = setting$sigma,
         weight = setting$weight,
         deriv = setting$deriv)
}, .parallel = TRUE)
# Extract coefficients
Betas <- ldply(modelfits, function(mod) data.frame(beta = mod$beta, grid = seq.int(365)/365))
# Plot
pF <- ggplot(data = Betas, aes(x = grid, y = beta)) + geom_line()
pF <- pF + facet_grid(weight ~ sigma, labeller = label_both)
pF + scale_x_continuous(breaks = seq(0, 1, length = 12),
                        labels = monthLetters) +
                  xlab("Calendar time") + ylab(expression(beta(m))) +
                  coord_flip()
```

Figure 2 shows the result for various values of the tuning constants. Particularly, $\sigma$ (`sigma` on Figure 2) corresponds to $\lambda/\sqrt{2\log p}$ in James et al. (2009), where $p$ is the dimension of the chosen basis ($p = 365$ here). The larger $\sigma$ is the sparser the solution is. Furthermore, $w$ (`weight` on Figure 2) is the weight that is placed on the zeroth derivative relative to the other derivative. So, if $w = 0$ the zeroth derivative is ignored.

As is apparent in Figure 2, for all values of the tuning constants, FLiRTI identifies that the temperatures roughly between October and November are important in determining the log precipitation, and particularly that the lower the temperature at those months the higher is the total annual precipitation. Note also that for low values of $\sigma$ ($\sigma = 0.002$ in Figure 2), an important effect of the temperature in the period roughly from February to May is also revealed, and, particularly, the larger the temperature at those months is, the higher is the total annual precipitation.

## 2.2 Multi-resolution elastic net results

We now estimate model (1) using the multi-resolution elastic net approach that is described in the main text. For that we replace the integral in model (1) with the discretized version

$$\sum_{g=1}^{G} \beta_g \bar{T}_{ig}, \qquad (2)$$

where $\bar{T}_{ig}$ is the average of the temperatures in the $g$th time interval for the $i$th weather station. That average is $\bar{T}_{ig} = \sum_{m \in M_g} T_i(m)/n_g$ where $M_g = (365(g-1)/G, 365g/G)$ and $n_g$ is the

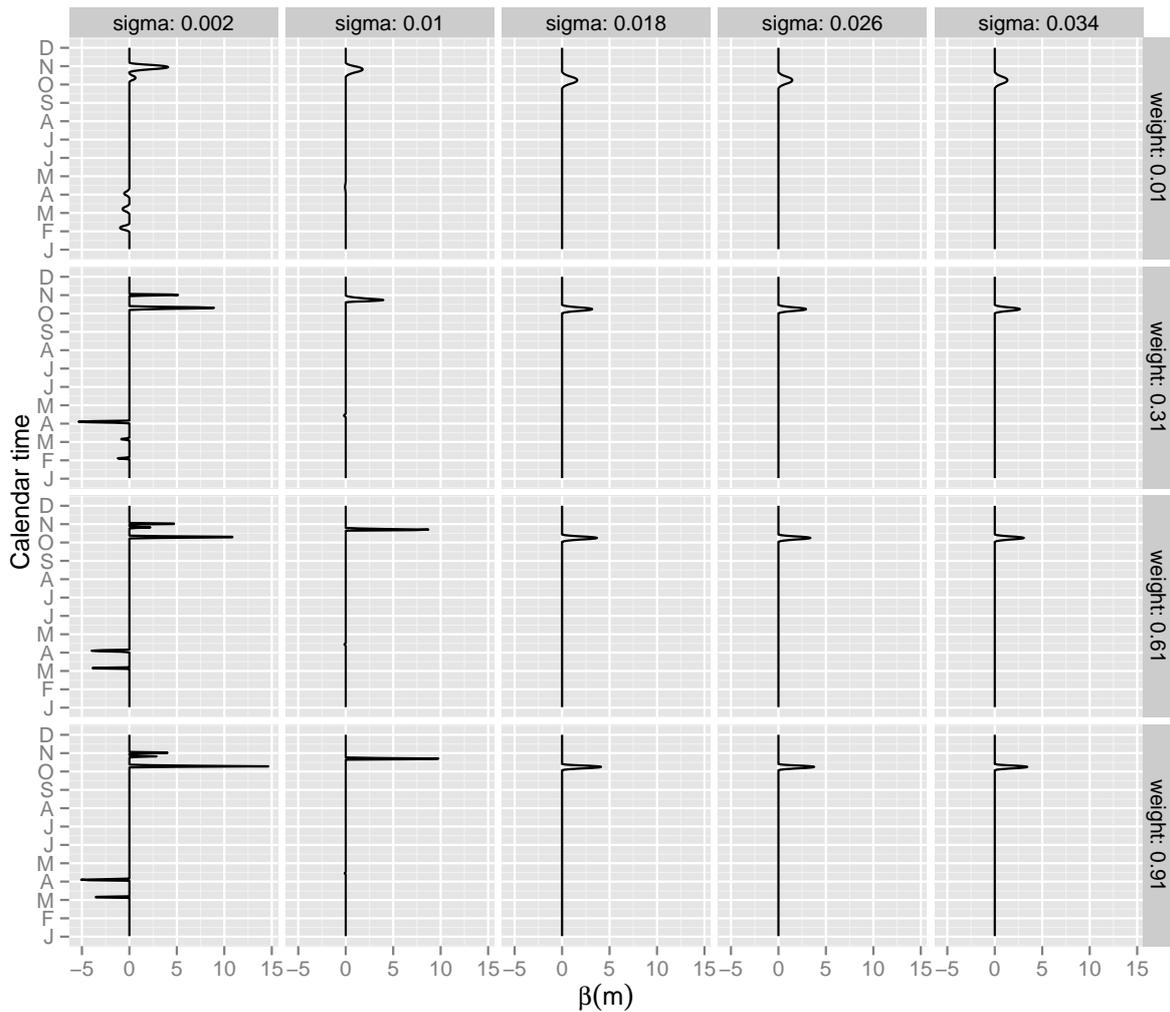

Figure 2: Estimates of $\beta(m)$ in model (1) for various values of the tuning constants $\sigma$ and $w$ in the FLiRTI approach. For $\beta(m)$, a simple grid basis is used that has dimension equal to the number of points that the value of the temperature is recorded at. Furthermore, $\beta(m)$ and the third-derivative of it are restricted in order to get a sparse piece-wise quadratic estimate.

number of observed temperatures in $M_g$. Figure 3 shows the results for $G \in \{40, 80, \ldots, 360\}$ and various settings for the tuning constants of the elastic net. Here, the elastic net is parameterized in terms of the $\lambda$ (`lambda` in Figure 3) and $s$ (`fraction` in Figure 3). The tuning constant $\lambda$ controls the weights that is assigned in the L2 norm in the elastic net, with $\lambda = 0$ corresponding to the LASSO fit. The constant $s$ is the fraction of the L1 norm and the smaller it is the sparser is the solution. For more details on those tuning constants, see Zou and Hastie (2005, Section 3.5).

As is the case for FLiRTI, for all values of the tuning constants considered in Figure 3, multi-resolution elastic net identifies that the temperatures roughly between October and November are important in determining the log precipitation, and particularly that the higher the temperature at those months the higher is the total annual precipitation (note that the sign of the non-zero estimates for $\beta_g$ is positive across resolutions). For larger values of the fraction (i.e. as the sparseness of the solution is reduced), multi-resolution elastic net also identifies that the temperature in the months February to March and April to June has a negative effect on the logarithm of the total annual precipitation. These results agree with the results from FLiRTI.

In the main text cross-validation is used for the selection of the tuning constants for the

elastic net. That process would be equivalent to selecting the best combination of $\lambda$ and $s$ for each resolution in Figure 3 according to a cross-validation criterion. Then the optimal resolution had been selected based on the squared prediction error on a test set.

Finally, note that the analysis of the setting in the main text could not have been performed directly with the FLiRTI approach, because that setting involves one functional and 10 scalar covariates. An appropriate extension of the FLiRTI approach would require inducing sparseness to the coefficients of the chosen basis expansion for the functional parameter and the parameters of the scalar covariates simultaneously.

```r
fitEN <- function(nbreaks = 101, Y,
                  Xf, # functional covariates
                  Xs = NULL, # scalar covariates
                  fraction = 0.1,
                  lambda = 0) {
    grid <- seq.int(ncol(Xf))/ncol(Xf)
    labs <- cut(grid, breaks = seq(0, 1, length = nbreaks))
    Xfagg <- t(apply(Xf, 1, function(tt) tapply(tt, labs, mean)))
    gridG <- tapply(grid, labs, function(x) mean(range(x)))
    X <- cbind(Xfagg, Xs)
    fitMod <- enet(X, Y, lambda = lambda)
    finalCoefs <- predict(fitMod, s = fraction,
                          type = "coefficients", mode = "fraction")$coef
    list(coefficients = finalCoefs, grid = gridG, lambda = lambda,
         fraction = fraction)
}
# Settings for the tuning constants
settingsM <- expand.grid(resolution = seq(40, 360, 40),
                         fraction =  c(0.001, 0.05, 0.1, 0.15, 0.2),
                         lambda = c(0.01, 0.02, 0.2, 1))
# Fits using multi-resolution elastic net
fits <- alply(settingsM, 1, function(setting) {
    fitEN(nbreaks = setting$resolution,
          Y = Y,
          Xf = Xf,
          fraction = setting$fraction,
          lambda =  setting$lambda)
}, .parallel = TRUE)
# Extract coefficients
coefs <- ldply(fits, function(fit) {
    with(fit, cbind(coefs = coefficients, grid = grid))
})
coefs$sign <- ifelse(coefs$coefs > 0, "+", "-")
# Plot
pM <- ggplot(data = coefs) +
    geom_point(aes(x = resolution, y = grid), shape = "-", colour = "grey")
pM <- pM +  geom_point(data = subset(coefs, coefs!=0),
                       aes(x = resolution, y = grid, shape = sign), size = 0.9)
pM + scale_shape_manual(values=c(24, 25)) + ylab("Calendar time") +
    scale_y_continuous(breaks = seq(0, 1, length = 12),
                       labels = monthLetters) +
                          facet_grid(lambda ~ fraction, labeller = label_both)
```

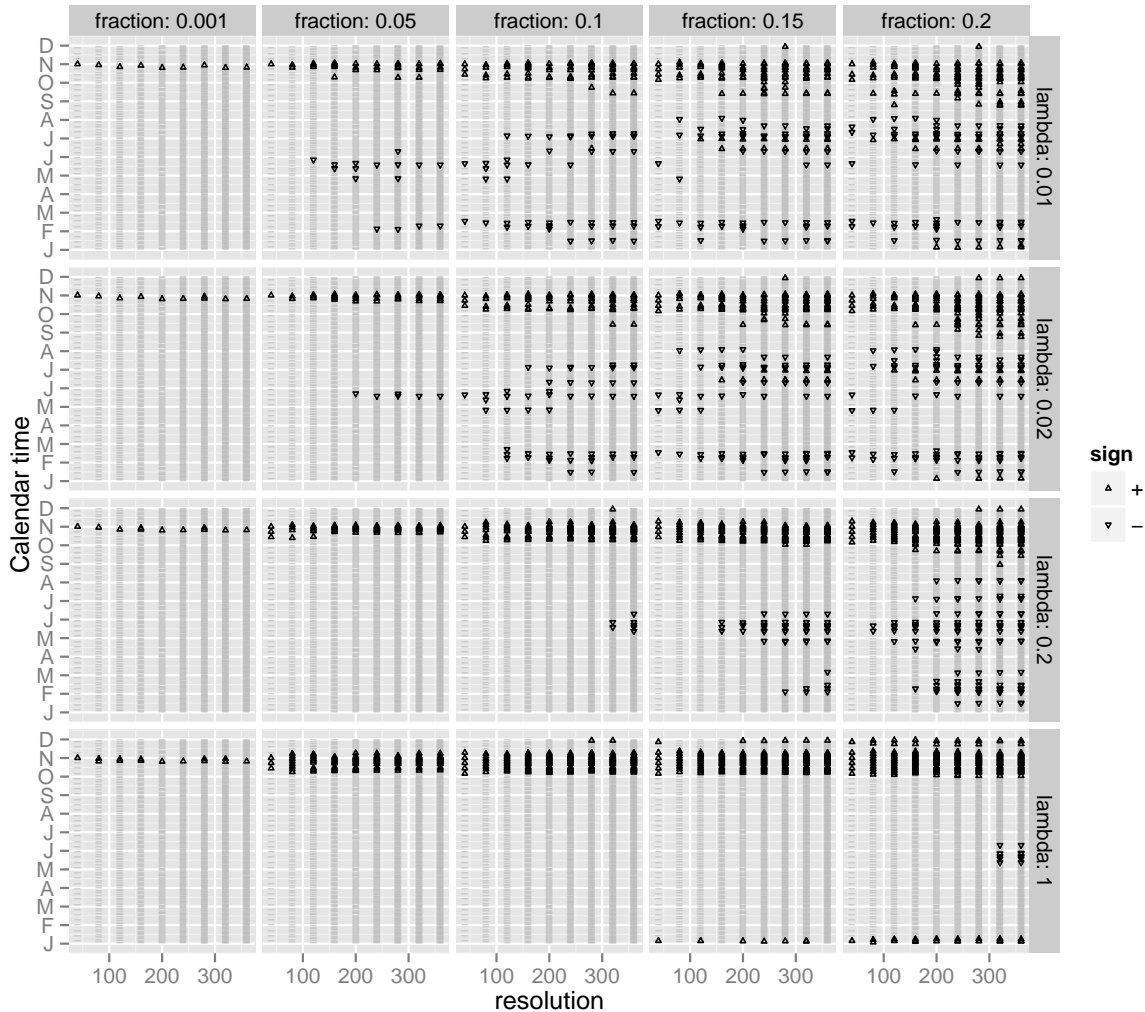

Figure 3: Estimates of $\beta_g$ in (2) model (1) for various values of the tuning constants $\lambda$ and $s$ of the multi-resolution elastic net and across resolutions. For each resolution, only the non-zero estimates are plotted and at the mid-points of the intervals $M_1, \ldots, M_G$. The plotting character depends on the sign of each estimate as shown in the legend.



\end{document}